\title{magnetic-map-TCLW-revision-8}
\documentclass[namedreferences]{solarphysics}

\usepackage[hyperref,optionalrh]{spr-sola-addons} 
\usepackage{graphicx}        
\usepackage[all]{hypcap} 
\usepackage{hyperref} 
\usepackage[all]{hypcap} 
\usepackage{amssymb}        
\usepackage{xcolor}           
\usepackage{multirow}
\usepackage{enumitem}
\usepackage{bm}
\usepackage[english]{isodate}
\usepackage{siunitx}
\usepackage[caption=false]{subfig}
\usepackage[labelformat=parens,labelsep=quad,skip=3pt]{caption}
\usepackage{breakurl}        

\chardef\us=`\_

\makeatletter
\newcommand\tsb[1]{\@textsubscript{\selectfont#1}}
\def\@textsubscript#1{{\m@th\ensuremath{_{\mbox{\fontsize\sf@size\z@#1}}}}}
\newcommand\tsp[1]{\@textsuperscript{\selectfont#1}}
\def\@textsuperscript#1{{\m@th\ensuremath{^{\mbox{\fontsize\sf@size\z@#1}}}}}

\begin{document}

\begin{article}
\begin{opening}

\title{Mapping Magnetic Field Lines for an Accelerating Solar Wind}

\author[addressref={aff1},corref,email={samira.tasnim@sydney.edu.au}]{\inits{S. }\fnm{S.}~\lnm{Tasnim}}
\author[addressref=aff1,email={iver.cairns@sydney.edu.au}]{\inits{I. H. }\fnm{Iver H.}~\lnm{Cairns}}
\author[addressref={aff1},email={bo.li@sydney.edu.au}]{\inits{B.}\fnm{B.}~\lnm{Li}}
\author[addressref={aff1},email={michael.wheatland@sydney.edu.au}]{\inits{M. S.}\fnm{M. S.}~\lnm{Wheatland}}

\address[id=aff1]{School of Physics, The University of Sydney, NSW 2006, Australia}

\runningauthor{Tasnim, Cairns, Li, \& Wheatland}
\runningtitle{Mapping Magnetic Field Lines}

\begin{abstract}
Mapping of magnetic field lines is important for studies of the solar wind and the sources and propagation of energetic particles between the Sun and observers. A recently developed mapping approach is generalised to use a more advanced solar wind model that includes the effects of solar wind acceleration, non-radial intrinsic magnetic fields and flows at the source surface/inner boundary, and conservation of angular momentum. The field lines are mapped by stepping along local magnetic field $\bm{B}$ and via a Runge-Kutta algorithm, leading to essentially identical maps. The new model's maps for Carrington rotation CR 1895 near solar minimum (19 April to 15 May 1995) and a solar rotation between CR 2145 and CR 2146 near solar maximum (14 January to 9 February 2014) are compared with the published maps for a constant solar wind model. The two maps are very similar on a large scale near both solar minimum and solar maximum, meaning that the field line orientations, winding angles, and connectivity generally agree very well. However, close inspection shows that the field lines have notable small-scale structural differences. An interpretation is that inclusion of the acceleration and intrinsic azimuthal velocity has significant effects on the local structure of the magnetic field lines. Interestingly, the field lines are more azimuthal for the accelerating solar wind model for both intervals. In addition, predictions for the pitch angle distributions (PADs) for suprathermal electrons agree at the $90$ -- $95\%$ level with observations for both solar wind models for both intervals.
\end{abstract}
\end{opening}

\section{Introduction}
     \label{S-Introduction} 
The heliospheric magnetic field structure and configuration are vital for understanding connections to the solar wind's source regions and the transport of nonthermal particles between the Sun and the Earth or even beyond \citep{Feldman1975,Rosenbauer1977,Owens2013}. Examples include suprathermal Strahl electrons with energy $\approx 80\, {\rm eV}-2 \,{\rm keV}$ \citep{Rosenbauer1977}, solar energetic particles (SEPs) with energy $\gsim 1 \,\rm MeV$ (e.g. relativistic electrons, protons and other ions) \citep{Kutchko1982,Richardson1991,Ruffolo2006}, and the subrelativistic to relativistic electrons that produce type III solar radio bursts \citep{Suzuki1985,Reiner1995}. These particles move much faster than the thermal plasma particles and the  $\bm{E}\times \bm{B}$ convection velocity, and thus primarily move along the magnetic field lines. The pitch angle distributions (PADs) of the strahl or heat flux electrons are commonly used to trace the magnetic field configurations \citep{Schatten1668,Owens2013,Li2016a,Li2016b}. 
     
The combination of solar rotation and solar wind outflow leads naturally to spiral-like magnetic field lines \citep{Parker1958}. Observation shows that the magnetic field lines of the solar wind are approximately Parker-like when averaged over many solar rotations \citep{Thomas1980,Burlaga1982,Bruno1997}. However, the magnetic field lines are often not Parker-like on timescales less than a solar rotation \citep{Forsyth1996,Borovsky2010,Hagen2011,Hagen2012,Tasnim2016,Tasnim2018}.
SEPs are associated with flares, coronal mass ejections (CMEs), and other solar activity associated with complex solar magnetic field structures \citep{Suzuki1985,Kutchko1982,Richardson1991,Ruffolo2006,Owens2013}. 

A number of solar wind models \citep{Smith1979,Fisk1996,Gosling2002,Schwadron2010,Hagen2011,Hagen2012,Tasnim2016,Tasnim2018} have been proposed to explain the deviation of the magnetic field lines from the Parker spiral. In this paper, we focus on the \citet{Tasnim2018} model which includes the effects of solar wind acceleration and intrinsic non-radial velocity and magnetic fields at the wind's effective source region (or inner boundary), and imposes conservation of angular momentum. Specifically, we update the mapping work of \citet{Li2016a} and \citet{Li2016b} to include the more advanced solar wind model of \citet{Tasnim2018}, rather than the constant wind speed models of \citet{Hagen2011,Hagen2012} and \citet{Tasnim2016}.  

Several approaches were developed previously to map the solar wind magnetic field on large scales from the Sun to the Earth \citep{Schatten1668,Nolte1973a,Nolte1973b,Li2016a,Li2016b}. \citet{Schatten1668} attempted first to map the ecliptic magnetic field lines on a large-scale using a wind model and near-Earth spacecraft magnetic field data. Their maps showed that field lines are usually more azimuthally oriented than the Parker spiral near solar maximum, but are more spiral-like and open near solar minimum. Later, \citet{Nolte1973a,Nolte1973b} proposed another approach, in which they estimated the high coronal source regions of solar wind plasma observed at the near-Earth location. However, these mapping approaches were unable to differentiate between open and closed field lines \citep{Gosling1974}. 

Recently, \cite{Li2016a} developed a mapping approach that steps along $\bm{B}({\bf r})$ (the $\bm{B}$-step algorithm) for fields obtained using a two-dimensional (2D) equatorial solar wind model \citep{Hagen2011,Hagen2012} and near-Earth (1 AU) data. This approach included an intrinsic non-radial magnetic field component at the photosphere and successfully mapped the field lines with more than 90\% agreement with the observed pitch angle distribution data \citep{Li2016a}. \citet{Li2016b} then applied the mapping approach to explore the correlation between the mapped field lines and the observed path of a type III solar radio burst, while \citet{Li2016c} showed that the model predicted well the occurrence of field line inversions. However, all these mapping approaches \citep{Schatten1668,Nolte1973a,Nolte1973b,Li2016a,Li2016b} assume that the radial solar wind outflow has a constant speed between the Sun and 1 AU. Moreover, the models neglect intrinsic non-radial velocities and magnetic fields at the source surface, conservation of angular momentum, as well as deviations from corotation at the source surface \citep{Schatten1668, Nolte1973a, Nolte1973b, Li2016a, Li2016b}. 

We employ the accelerating solar wind model of \citet{Tasnim2018} and near-Earth (1 AU) solar wind data for a solar rotation CR1895 (19 April to 15 May 1995) and a solar rotation between CR 2145 and CR 2146 (14 January to 9 February 2014) to predict the magnetic field vectors and to map the magnetic field lines between the Sun and 1 AU. Put in other words, we predict the plasma quantities (mass density, radial and azimuthal velocity), and magnetic field components (radial and azimuthal magnetic field) from the source surface / inner boundary, 
$r_{\rm s}$, to all $r$ using \citet{Tasnim2018}'s model and {\it Wind} spacecraft data at 1 AU. Note that this model only includes the intrinsic azimuthal velocity and magnetic field components, but can not explain the dynamical variations between the source surface and 1 AU. However, \citet{Hu1992,Hu1993,Odstrcil1994,Richardson1996} demonstrated the large values of observed azimuthal velocity at 1 AU and the non-Parker magnetic field lines can be due to dynamic effects as stream-stream interactions (SIRs) at corotating interaction regions (CIRs).

To assess the mapping approach, we develop and use a Runge-Kutta algorithm, as well as the $\bm{B}$-step method. Both algorithms predict nearly identical field lines and maps.  We then compare the field lines and map with those for the constant radial speed solar wind model of \citet{Hagen2011,Hagen2012}, and test the two models' predictions for the pitch angle distributions (PADs) of suprathermal electrons \citep{Rosenbauer1977,Gosling1987,Li2016a} observed for the same solar rotation periods. 

The paper is structured as follows. Section~\ref{Models} briefly describes the two solar wind models employed to predict ${\bf B}({\bf r})$. Section~\ref{mapping-approach} summarises the $\bm{B}$-step \citep{Li2016a,Li2016b} and Runge-Kutta mapping algorithms. Section~\ref{mapped-flines} assesses the magnetic field maps predicted for the two solar wind models using both mapping algorithms for the two solar rotation periods {\color{black} and describes the topology of the the magnetic field lines}. Section~\ref{PAD-classes} compares the PADs predicted for the two models and the observations. Section~\ref{dis-con} discusses the results and concludes the paper.

\section{Accelerating and Constant Radial Speed Solar Wind Models
and Predicted Magnetic Fields}
\label{Models}
First, we consider \citet{Tasnim2018}'s accelerating solar wind model and its predictions for ${\bf B}({\bf r})$ vectors. This equatorial plane model uses the time-steady isothermal equation of motion to describe the radial acceleration of the solar wind. It then combines the accelerating solar wind profile with Faraday's Law, Gauss's Law, and the MHD  equations for frozen-in field magnetic fields and angular momentum conservation. {\color{black} Note that \citet{Tasnim2018}'s model solves Parker’s isothermal wind model using an implicit method which considers the gravitational effects on the acceleration of the solar wind and assumes the magnetic field terms are unimportant for the acceleration. However, this model does not include dynamical effects which leads to an ignorance of acceleration by the stream-stream interactions \citep{Richardson2018}.}

The model assumes the solar wind's sources are constant over a solar rotation. This assumption allows a fixed global pattern to rotate with the Sun, with the Earth moving through this fixed pattern. The wind varies with heliocentric distance $r$ and heliolongitude $\phi$ in the rotating frame. We use primed ($\prime$) variables for the magnetic field, position, and plasma quantities in the corotating frame whereas unprimed variables are in the inertial frame. The magnetic field and velocity in the inertial frame are denoted $\bm{B}=(B_r,B_\phi)$ and $\bm{v}=(v_r,v_\phi)$, respectively. We assume $B_\phi^\prime(r^\prime,\phi^\prime) = B_\phi(r,\phi,t)$. We also consider the variations of $\bm{B}$ and $\bm{v}$ in $\phi$ are small locally and can be neglected in comparison with the variations with $r$ in Gauss's Law, {\it i.e.} $\partial B_\phi^\prime/\partial \phi^\prime \approx 0$. Then integrating Gauss's Law over $r$ leads to:
\begin{equation}
r^{\prime 2} B_r^\prime(r^\prime,\phi^\prime)=r^{2} B_r(r,\phi,t)=r_{\rm s}^{\prime 2} B_r^\prime(r_{\rm s}^\prime,\phi_{\rm s}^\prime),\label{Gauss-law}
\end{equation}
where the inner-boundary/source surface of the solar wind is described by $(r_{\rm s}^\prime, \phi_{\rm s}^\prime)$ in the rotating frame. {\color{black} Note that recent investigation of \citet{Dosa2017} found the existence of longitudinal variations of the magnetic flux density and that the radial magnetic field does not necessarily fall with $r^2$. Therefore, a future improvement of the \citet{Tasnim2018}'s solar wind model will address and fit such radial profile, and investigate the implications on the magnetic field line mapping.}

Considering time-steady frozen-in flow in the corotating frame leads to  
\begin{eqnarray}
&& \hspace{-2ex} B_\phi^\prime(r^\prime,\phi^\prime)=B_\phi(r,\phi,t)= \frac{r_{\rm s} v_r(r_{\rm s},\phi_{\rm s},t)B_\phi(r_{\rm s},\phi_{\rm s},t) }{r v_r(r,\phi,t)}-\frac{r_{\rm s} \delta v_\phi(r_{\rm s},\phi_{\rm s},t)B_r(r_{\rm s},\phi_{\rm s},t)}{r v_r(r,\phi,t)}\nonumber\\
&& \hspace{+12ex} +\frac{v_\phi(r,\phi,t)B_r(r,\phi,t)}{v_r(r,\phi,t)}-\frac{\Omega r_{\rm s}^2 B_r(r_{\rm s},\phi_{\rm s},t)}{r v_r(r,\phi,t)},\label{TCW-bphi}
\end{eqnarray}
where $\Omega=2\pi/T$ is the Sun's rotation frequency and $T = 27\,{\rm days}= 648\, {\rm hours}$ is the Sun's synodic rotation period. Here $v_\phi^\prime(r_{\rm s}^\prime,\phi_{\rm \phi}^\prime)=\delta v_\phi^\prime(r_{\rm s}^\prime,\phi_{\rm \phi}^\prime)=\delta v_\phi(r_{\rm s},\phi_{\rm s},t)$ is the deviation of the azimuthal speed from corotation at the source surface. It is not assumed that ${\bf v}$ and ${\bf B}$ are parallel anywhere, so that in general a convection electric field ${\bf E} = - {\bf v} \times {\bf B} \ne 0$ exists at all ${\bf r}$. More detailed justifications of these expressions are given in \citet{Tasnim2016} and \citet{Tasnim2018}.

Second, we consider the \citet{Hagen2012} model used by \citet{Li2016a,Li2016b,Li2016c} to predict $\bm{B}(\bm{r})$. This model assumes $v_r(r,\phi,t)$ is constant along streamlines, but includes a non-zero intrinsic azimuthal magnetic field at the source surface/inner-boundary. It assumes corotation at the inner boundary to obtain $v_{\phi}(r_{s},\phi_{s})$ and does not conserve angular momentum. The predictions are Equation~\ref{Gauss-law} and
$B_\phi(r,\phi,t)$ for the constant radial speed solar wind model is
\begin{eqnarray}
B_\phi(r,\phi,t)= \frac{B_\phi(r_{\rm s},\phi_{\rm s},t) r_{\rm s}}{r}-\frac{B_r(r,\phi,t)\Omega (r-r_{\rm s})}{v_r(\phi,t)},\label{Bphi-const}
\end{eqnarray}
where we use $B_\phi^\prime(r^\prime,\phi^\prime)=B_\phi(r,\phi,t)$ for consistency with Equation \ref{TCW-bphi}.

We fit 1 AU data from the {\it Wind} spacecraft to the two solar wind models to obtain ${\bf B}({\bf r})$ for a particular Carrington rotation. First, we extract the magnetic field components $B_\phi^\prime(r_{\rm s}^\prime,\phi_{\rm s}^\prime)$ and  $B_r^\prime(r_{\rm s}^\prime, \phi_{\rm s}^\prime)$, intrinsic non-radial velocity component $\delta v_\phi(r_{\rm s}^\prime,\phi_{\rm s}^\prime)$, and accelerating radial wind speed $v_r(r,\phi)$ for an accelerating solar wind model using 1 AU data -- mathematical expressions and detailed descriptions are available in \citet{Tasnim2018}. Then, we predict $B_r^\prime(r^\prime,\phi^\prime)=B_r(r,\phi,t)$ and $B_\phi^\prime(r^\prime,\phi^\prime)=B_\phi(r,\phi,t)$ for the accelerating solar wind model using Equations \ref{Gauss-law} and \ref{TCW-bphi} using the extracted $B_\phi^\prime(r_{\rm s}^\prime,\phi_{\rm s}^\prime)$, $B_r^\prime(r_{\rm s}^\prime, \phi_{\rm s}^\prime)$, and $\delta v_\phi(r_{\rm s}^\prime,\phi_{\rm s}^\prime)$. Similarly, we predict $B_\phi(r_{\rm s},\phi_{\rm s},t)$ and $B_\phi(r,\phi,t)$ for the constant solar wind model \citep{Hagen2012}. The predicted values $B_\phi^\prime$ and $B_r^\prime$ allow us to predict $\bm{B}(r,\phi,t)=\bm{B}^\prime(r^\prime,\phi^\prime)$ at any location $(r,\phi)$ in the equatorial plane for the corresponding solar wind model, {\it i.e.} application of
Equation \ref{TCW-bphi} yields $\bm{B}(r,\phi,t)$ for the accelerating solar wind model and Equation \ref{Bphi-const} gives  $\bm{B}(r,\phi,t)$ for the constant solar wind model.

\section{Mapping Algorithms: B-step and Runge-Kutta Approaches}
\label{mapping-approach}

The $\bm{B}$-step algorithm for mapping the field lines  \citep{Li2016a,Li2016b} simply  steps along $\bm{B}$ and $-\bm{B}$ from a starting point  $\bm{r}_0=(x_0,y_0)$ between the Sun and the Earth in the solar equatorial plane. 
Specifically, starting from $\bm{r}_0$, the field line is calculated by stepping along the locally--averaged $\bm{B}$ toward a new point $\bm{r}=(x,y)$ defined by 
\begin{eqnarray}
x & =  x_{0} + \Delta x & =  x_{0} + \sum_{i=1}^4 \frac{{B_{xi}}}{B_i} dl \ , \label{Bo_map_1} \\
y & =  y_{0} + \Delta y & =   y_{0}  + \sum_{i=1}^4 \frac{{B_{yi}}}{B_i} dl\ . \label{Bo_map_2}
\end{eqnarray}
Here $B_{xi}$ and $B_{yi}$ are the x and y components of the magnetic field vectors $\bm{B}_i(\bm{r}_i)$ with $i = 1,...,\,{\rm and}\,4$ for the four nearest grid points ${\bf r}_{i}$ to $\bm{r_0}$ that form a two dimensional cell that enclose the starting point $\bm{r_0}$ and $dl$ is the length step. This process continues until the chosen final heliocentric distance $\bm{r}$ is reached. The field lines are mapped in the opposite direction by the transformations $(\Delta x,\Delta y) \rightarrow (-\Delta x,-\Delta y)$ and $dl \rightarrow -dl$ in Equations \ref{Bo_map_1} - \ref{Bo_map_2}. Note that these equations employ a local average of $\bm{B}$, but do not use interpolation. This should be more stable, overall, but will not in general preserve the direction of $\bm{B}$ at 1 AU or exactly satisfy Gauss's Law.

\begin{figure}[!tbp]
  \centering
  \begin{minipage}[a]{0.77\textwidth}
    \includegraphics[width=\textwidth,height=0.42\textheight]{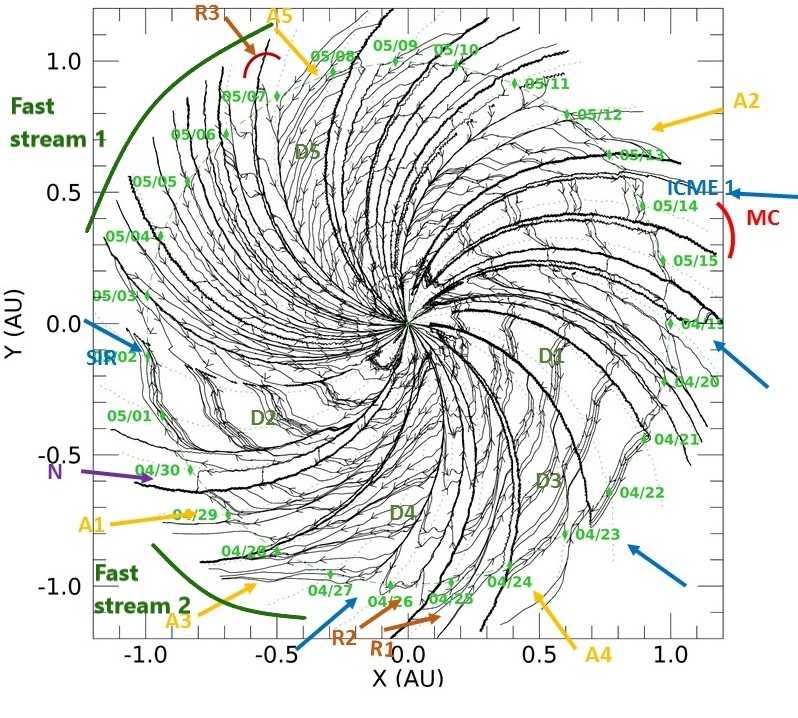}
  \end{minipage}
  \hfill
  \begin{minipage}[a]{0.7\textwidth}\hspace*{-1.5em}
    \includegraphics[width=\textwidth]{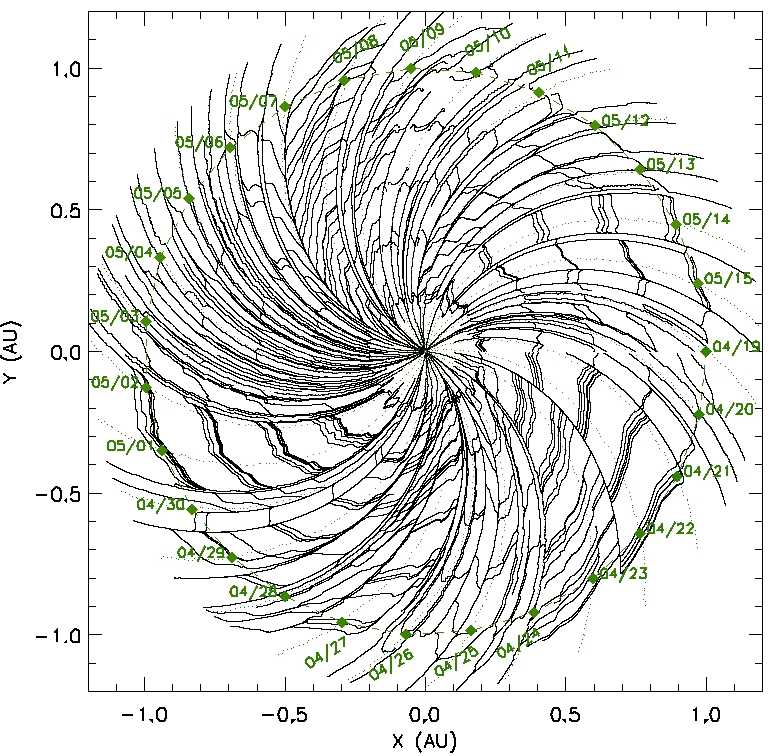}
    \caption{Maps of magnetic field lines predicted in the equatorial plane using the ({\it top}) $\bm{B}$-step and ({\it bottom}) Runge-Kutta approaches for the accelerating solar wind model of \citet{Tasnim2018} for the solar rotation period 19 April to 15 May 1995 (CR 1895). The reference frame is fixed on the Sun (the primed frame) and the Earth moves clockwise around the Sun with increasing time. {\it Green diamond markers} show Earth's location at 00 UT on each specified date while {\it arrows} show the direction of ${\bf B}({\bf r})$. Symbols MC identify magnetic clouds, ${\rm A}i$ for $i=1,2,3...$ periods with azimuthal field lines, and likewise ${\rm D}i$ periods with low $B$, ${\rm R}i$ radial field lines, and $\rm N$ field lines in the anticlockwise direction. 
}
    \label{mapping_1}
  \end{minipage}
\end{figure}

\begin{figure}[!tbp]
  \centering
  \begin{minipage}[a]{0.77\textwidth}
    \includegraphics[width=\textwidth]{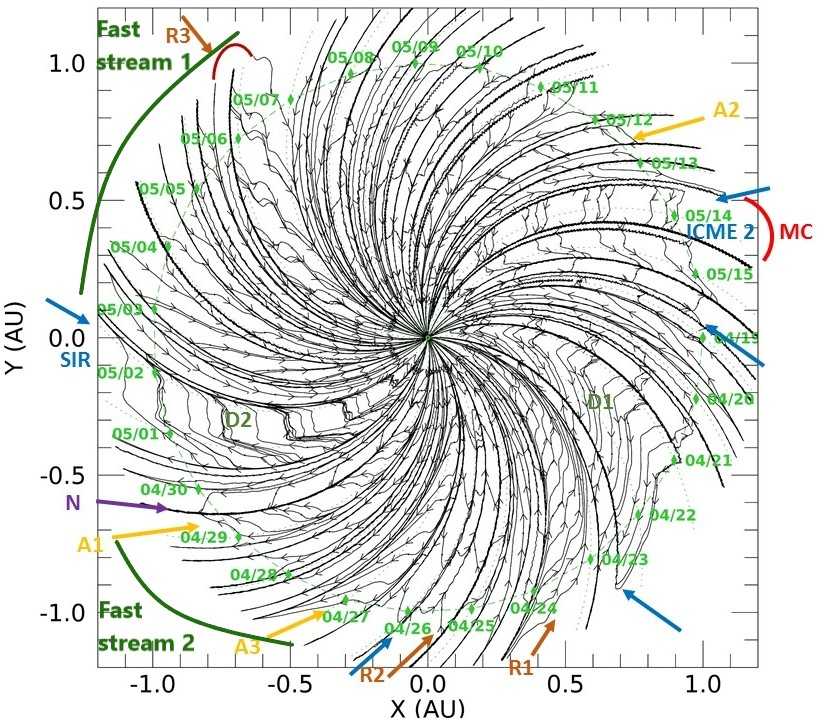}
  \end{minipage}
  \hfill
  \begin{minipage}[a]{0.7\textwidth}\hspace*{-1.2em}
    \includegraphics[width=\textwidth]{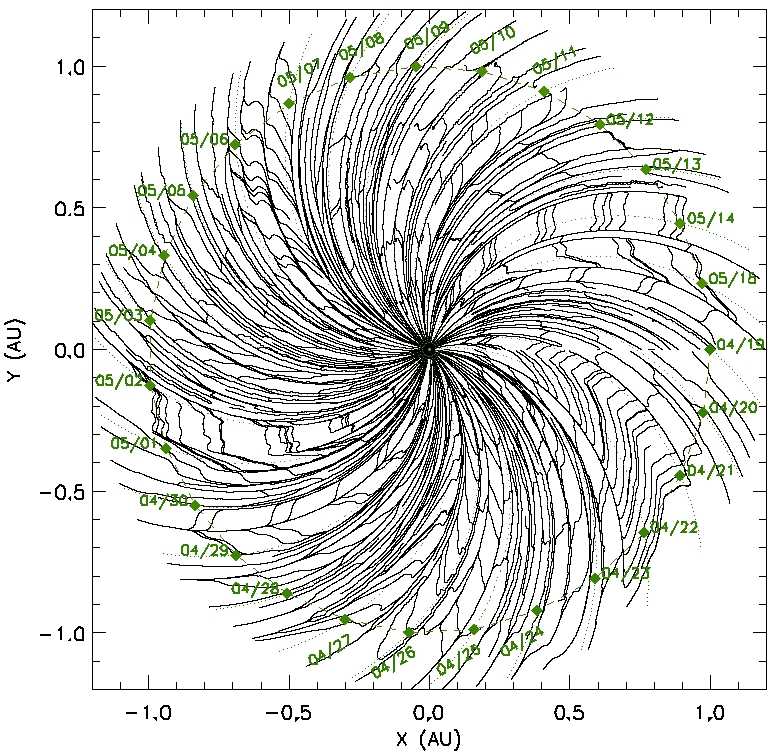}
    \caption{As for Figure \ref{mapping_1} but for the constant solar wind model of \citet{Hagen2011,Hagen2012}: ({\it top})  $\bm{B}$-step and ({\it bottom}) Runge-Kutta approaches for solar rotation CR 1895.}
    \label{mapping_2}
  \end{minipage}
\end{figure}


The new Runge-Kutta algorithm maps the magnetic field lines on a regular grid. We first linearly interpolate the components $B_r(r,\phi)$ and $B_\phi(r,\phi)$ onto a regular $(r,\phi)$ from those provided by the two solar wind models (regular in $r$ but not $\phi$). The regular $(r,\phi)$ grid has $r_{\rm reg}$ uniformly distributed from $R_\odot$ to $1.2$ AU and $\phi_{\rm reg}$ uniformly distributed from $0$ to $2\pi$. The magnetic field lines are defined in $(r_{\rm reg},\phi_{\rm reg})$ coordinates by 
\begin{equation}
\frac{B_\phi(r_{\rm reg},\phi_{\rm reg})}{d \phi}=\frac{r B_r(r_{\rm reg},\phi_{\rm reg})}{dr}.\label{runge-kutta}
\end{equation}
We trace the field lines by integrating Equation \ref{runge-kutta} using the fourth order Runge-Kutta method. 

We now consider the magnetic field maps predicted using these two algorithms for the accelerating solar wind model using Equations \ref{Gauss-law} and \ref{TCW-bphi} \citep{Tasnim2018} and for the constant speed solar wind model using Equations \ref{Gauss-law} and \ref{Bphi-const}  \citep{Hagen2011,Hagen2012}. Figure~\ref{mapping_1} compares magnetic field lines predicted for the accelerating model using the two algorithms, and Figure~\ref{mapping_2} compares the field lines using the above algorithms for the constant radial speed model.

By comparing the top and bottom panels within Figures~\ref{mapping_1} and ~\ref{mapping_2}, it is evident that the $\bm{B}$-step and Runge-Kutta algorithms predict almost identical field lines with only minor differences.  These strong similarities demonstrate the validity of the mapping algorithms. Put in other words, we can confidently use either algorithm for the same set of starting points to yield essentially the same map.

An important aspect of Figures~\ref{mapping_1}  and ~\ref{mapping_2} is that we have improved the displays to provide unbiased and ``global" maps that use a regular, array of starting points rather than biasing the display to those field lines that pass through a regular array of points at 1 AU. This is an important improvement of the displays of \citet{Li2016a, Li2016b, Li2016c}, which show field lines biased to starting points near $1$~ AU instead of showing unbiased global maps.

Although the two mapping algorithms yield nearly identical field lines, a few differences are present between the maps in Figures \ref{mapping_1} and \ref{mapping_2} for these algorithms; for example, consider the field lines between 23 April and 24 April. {\color{black} To be more precise, a field line near 23 April using Runge-Kutta algorithm is not connected back to the other field lines as with the $\bm{B}$-step algorithm.  Another field line close to 24 April using Runge-Kutta has not reached as close to other field lines as using $\bm{B}$-step algorithm. } One reason for these differences is that the $\bm{B}$-step algorithm uses 4-point averaged data from the ${\bf B}({\bf r})$ model while the Runge-Kutta algorithm linearly interpolates the model data. Therefore, neither algorithm preserves the model directions exactly and small differences are expected between the field lines predicted using the $\bm{B}$-step and Runge-Kutta algorithms. 

Another cause of differences is that the two algorithms treat field lines differently at locations corresponding to the start and the end of the solar rotation period, which border each other spatially but differ in time. In detail, the $\bm{B}$-step algorithm does not implement a spatial boundary between the field lines
and instead connects them by default \citep{Li2016a, Li2016b}, while the Runge-Kutta approach does not artificially connect these field lines. For example, the field lines between the labels 19 April and 15 May in Figures~\ref{mapping_1} (bottom) and ~\ref{mapping_2} (bottom) have discontinuities for the Runge-Kutta algorithm but are connected in the maps for the $\bm{B}$-step algorithm [{\it cf.} Figures~\ref{mapping_1} (top) and ~\ref{mapping_2} (top)].

Both approaches share some other limitations, including not recognizing sector boundaries (SBs) when moving along a field line and so wrongly connecting field lines across SBs. In addition, both algorithms occasionally become ineffective when significant variations of the local magnetic fields result in $\bm{B}\approx 0$. This condition stops the magnetic field lines from being mapped; {\it e.g.}, the field lines between 2 May and 3 May. 

\section{Magnetic Field Line Maps for the Accelerating and the Constant Radial Speed Wind}
\label{mapped-flines}
This section describes the detailed structures and orientations of the magnetic field lines for the accelerating and constant solar wind models for the solar rotation period between 19 April and 15 May 1995 (CR 1895) near solar minimum and the solar rotation period between 14 January and 9 February 2014 (bridging  CR 2145 and CR 2146) near solar maximum. We choose these periods since \citet{Li2016a} previously analysed the corresponding PAD data and compared them with the predictions for the constant solar wind model.

\subsection{Predicted Magnetic Maps for 19 April to 15 May 1995}
The detailed maps for CR 1895 are shown in Figures~\ref{mapping_1} and ~\ref{mapping_2},
for the accelerating and constant speed solar wind models, respectively. Clearly the two magnetic maps are almost identical on a broad scale. However, a closer view shows that some differences are present on small scales of order several days. Qualitative comparisons are as follows:

\begin{enumerate}[label = \roman{*})]

\item \textbf{Azimuthal field lines:} The field lines are azimuthal in the periods 29 -- 30 April, 27 -- 28 April, and 12 -- 13 May (labeled as A1, A2, and A3) for both models, but the field lines are more azimuthal for the accelerating model. {\color{black} Since A1, A2, and A3 are present in the maps for both of the models, an interpretation is that these azimuthal field lines demonstrate the contribution of the intrinsic azimuthal magnetic field. However, these lines are slightly more azimuthal for \citet{Tasnim2018}'s model due the inclusion of intrinsic azimuthal velocities. } Additional azimuthally oriented field lines, marked by A4 and A5 in the top panel of Figure~\ref{mapping_1}, demonstrate the effect of intrinsic azimuthal velocities and angular momentum conservation in the accelerating wind model.  

\item \textbf{Field line densities:} The field line densities are not always the same, corresponding to different field {\color{black} typologies}. Both Figures~\ref{mapping_1} and ~\ref{mapping_2} show lowly populated areas from 21 to 23 April and 1 to 2 May (labeled as D1 and D2). These lowly populated regions mostly correspond to strongly azimuthal field lines not just near $1$~AU but also close to the Sun. This result demonstrates the importance of including $B_\phi(r_{\rm s},\phi_{\rm s})$ in both solar wind models.

\item \textbf{Effects of accelerating $v_r(r,\phi,t)$ and $\delta v_\phi(r_{\rm s},\phi_s,t)$:} The maps in Figure ~\ref{mapping_1} show more lowly populated regions than Figure~\ref{mapping_2}, {\it e.g.}, from 23 to 24 May, 27 to 28 May, and 7 to 8 May (labeled as D3, D4, and D5). One interpretation is in terms of the combined effects of intrinsic azimuthal velocities $\delta v_\phi(r_{\rm s},\phi_{\rm s})\neq 0$, the accelerating radial wind speed $v_r(r,\phi,t)$, and conservation of angular momentum. Equation \ref{TCW-bphi} for the accelerating model shows that at small $r$ a lower $v_r(r,\phi,t)$ results in a larger $B_\phi(r,\phi,t)$. Similarly, a large deviation  $\delta v_\phi(r_{\rm s},\phi_{\rm s},t)$ from corotation in the anticlockwise direction results in a larger $B_\phi(r,\phi,t)$ at small $r$, based on  Equation \ref{TCW-bphi}, and  so more azimuthally oriented field lines. In contrast, the constant speed model does not allow these changes since it assumes strong corotation with the Sun $(\delta v_\phi =0)$ and a constant speed $(v_{r}=v_r(\phi))$ along streamlines.

\end{enumerate}
\begin{figure*}[htp]
   \subfloat{\label{rev}
      \includegraphics[width=.50\textwidth]{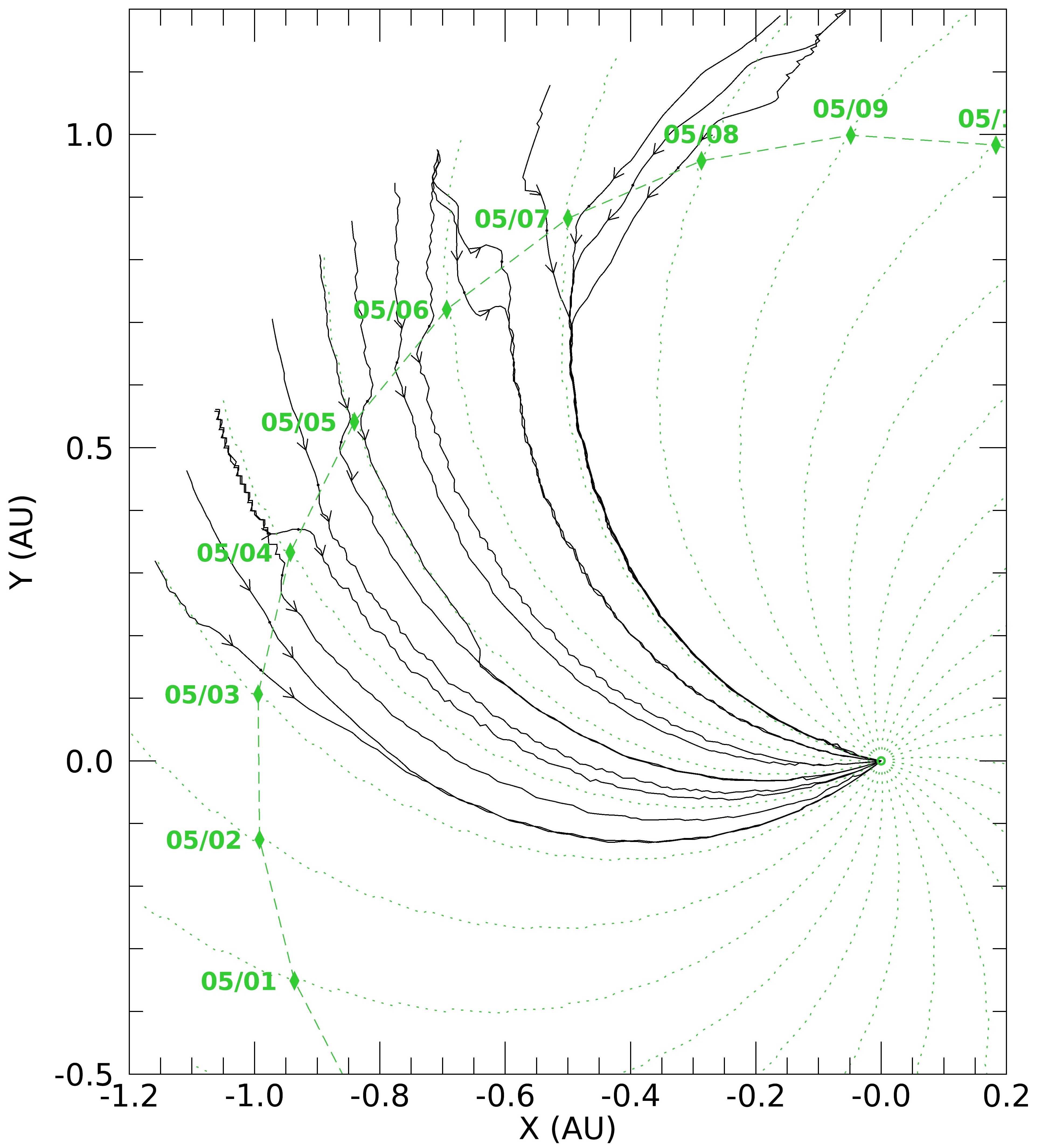}}
   \subfloat{\label{rev_sol}\hspace*{-0.9em}
      \includegraphics[width=.50\textwidth]{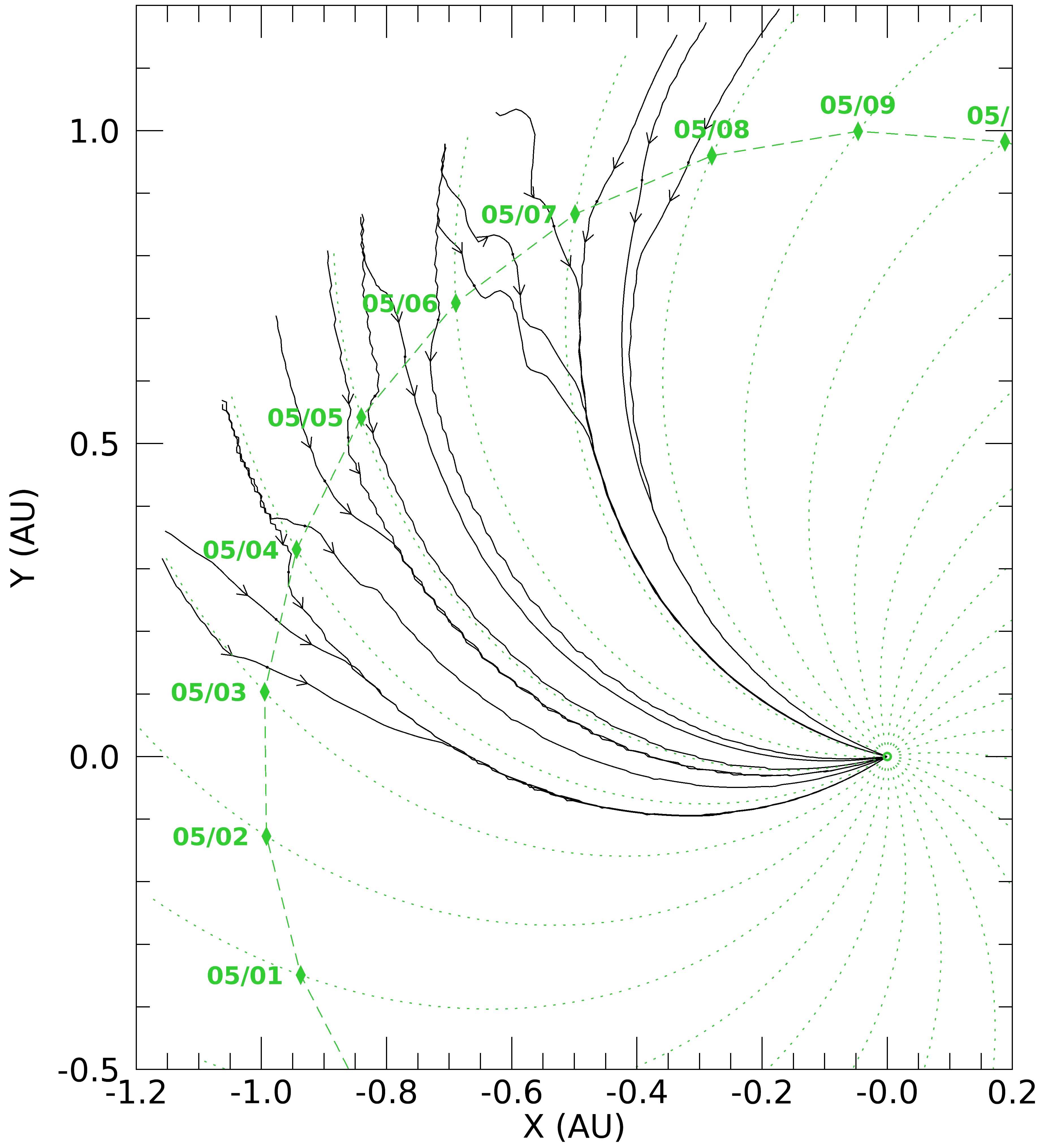}} 
\vspace*{-0.8em}
\subfloat{\label{rev-1}
      \includegraphics[width=.50\textwidth]{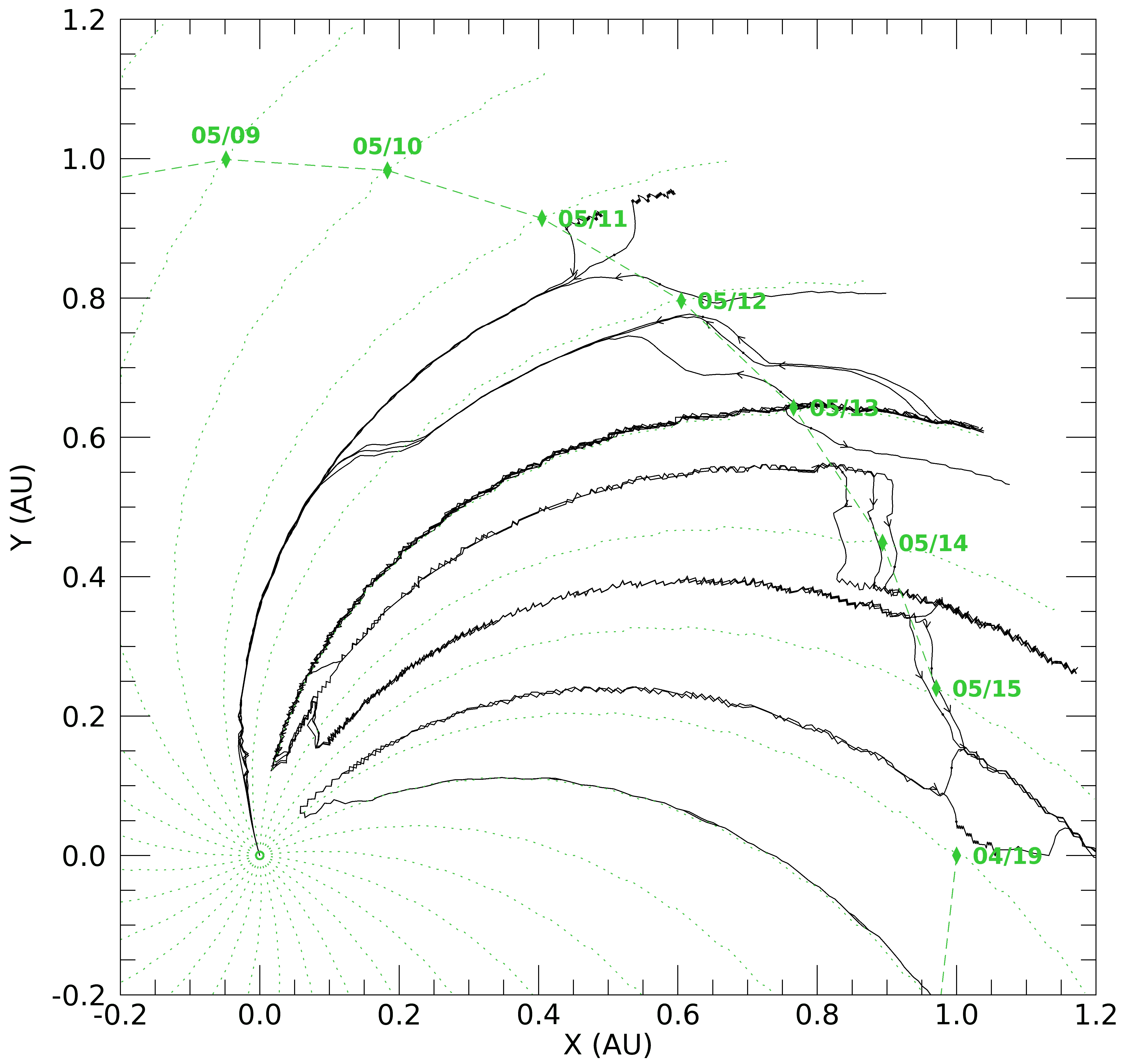}}
   \subfloat{\label{rev_sol-1}\hspace*{-0.8em}
      \includegraphics[width=.50\textwidth]{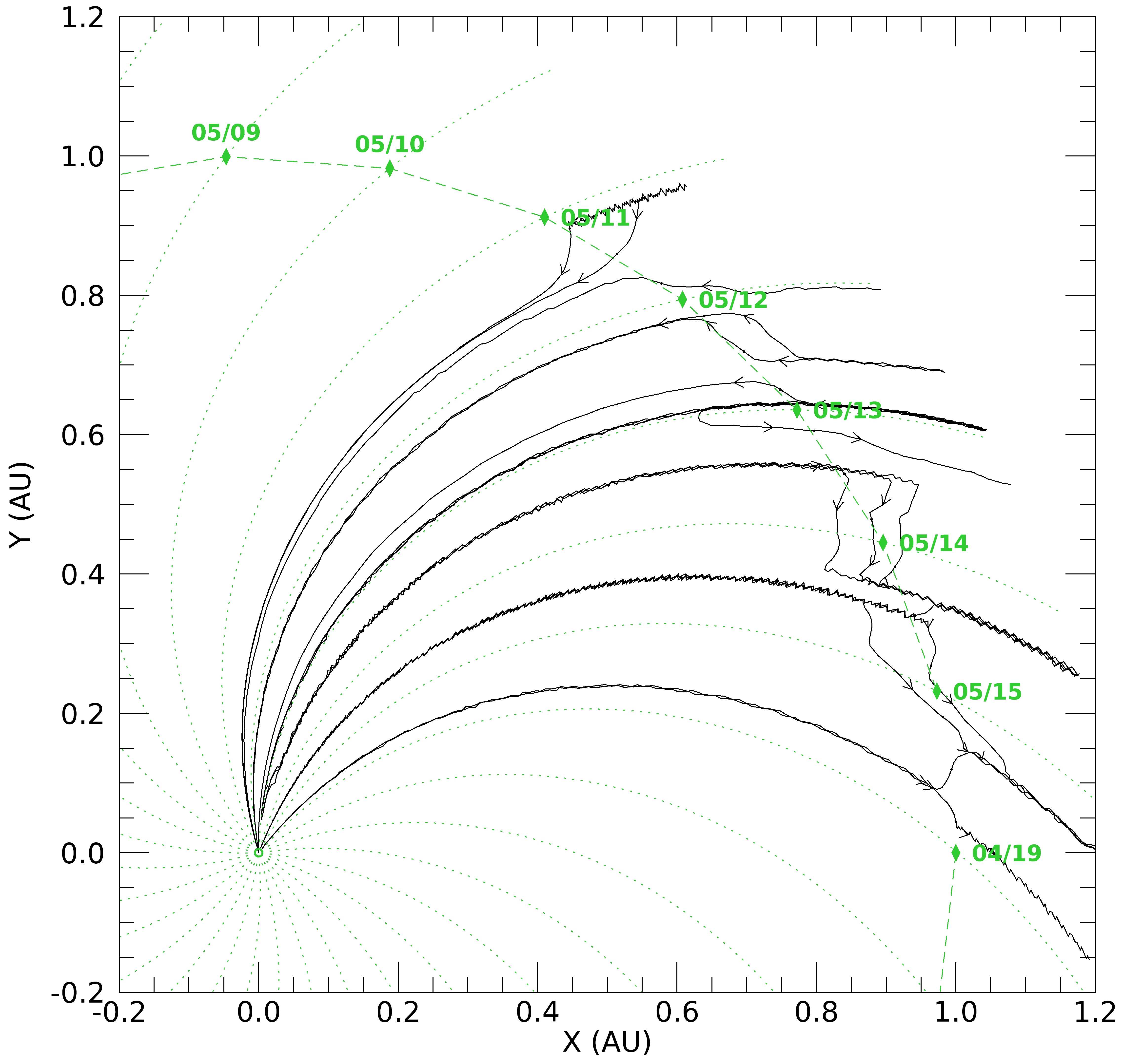}} 
   \caption{{\it Top panels} show the predicted magnetic field lines between 3 to 8 May 1995 for the accelerating ({\it left}) and the constant ({\it right}) wind models. Similarly, {\it bottom panels} show the lines between 11 to 19 May 1995 where the {\it left} and {\it right panels} present the field lines for the accelerating and the constant wind models, respectively. {\it Diamonds} and {\it arrows} are the same as Figures~\ref{mapping_1} and ~\ref{mapping_2}.}\label{mapping_3}
\end{figure*}

In addition to the above points, both Figures~\ref{mapping_1} and ~\ref{mapping_2} show some sudden changes in the magnetic field structure when CMEs and SIR events occur. (CMEs are not modelled in either model since both assume a constant pattern for the solar rotation. Note that both of these models are data driven model, so abrupt changes due to CMEs present in the predicted outputs. However, transient events (e.g. CMEs) are not modelled here.) For instance, the figures show open field lines from 13 to 14 May during the ICME event {\color{black}(marked as ICME 1 in Figure~\ref{mapping_1}) } \citep{Jian2010} and magnetic cloud MC \citep{Burlaga1982} (marked with a red curved line) whereas field line inversions occur during 2-3 May due to a stream interaction region (SIR). Blue arrows in Figures~\ref{mapping_1} and  ~\ref{mapping_2} indicate some additional local field inversions.

Figures~\ref{mapping_1} and ~\ref{mapping_2} also demonstrate that the models and $1$~AU data predict  approximately radial magnetic field lines at large $r$ (labeled R1, R2, R3), contrary to the nominal Parker spirals. The figures also show that both models and the $1$~AU data allow the field lines to orient in an anticlockwise direction instead of always having the  clockwise orientation of the nominal Parker spiral (labeled  N).

Figure~\ref{mapping_3} shows expanded views of the field lines predicted for 3 to 8 May (top two panels) and 11 to 19 May (bottom two panels) in 1995 for the two models. The field lines explicitly show notable path and connectivity differences for the two solar wind models  on small scales while being very similar on large scales. Again, the detailed dissimilarities in field-line connections indicate that the inclusion of wind acceleration, intrinsic non-radial velocities, and angular momentum  can significantly change the field line structures on small scales, even though their orientations at 1 AU remain very similar. 

\subsection{Magnetic Maps Near Solar Maximum: 14 January to 9 February 2014}
\label{solar-maximum}
\begin{figure}[!tbp]
  \centering
  \begin{minipage}[a]{0.71\textwidth}
    \includegraphics[width=\textwidth,height=0.42\textheight]{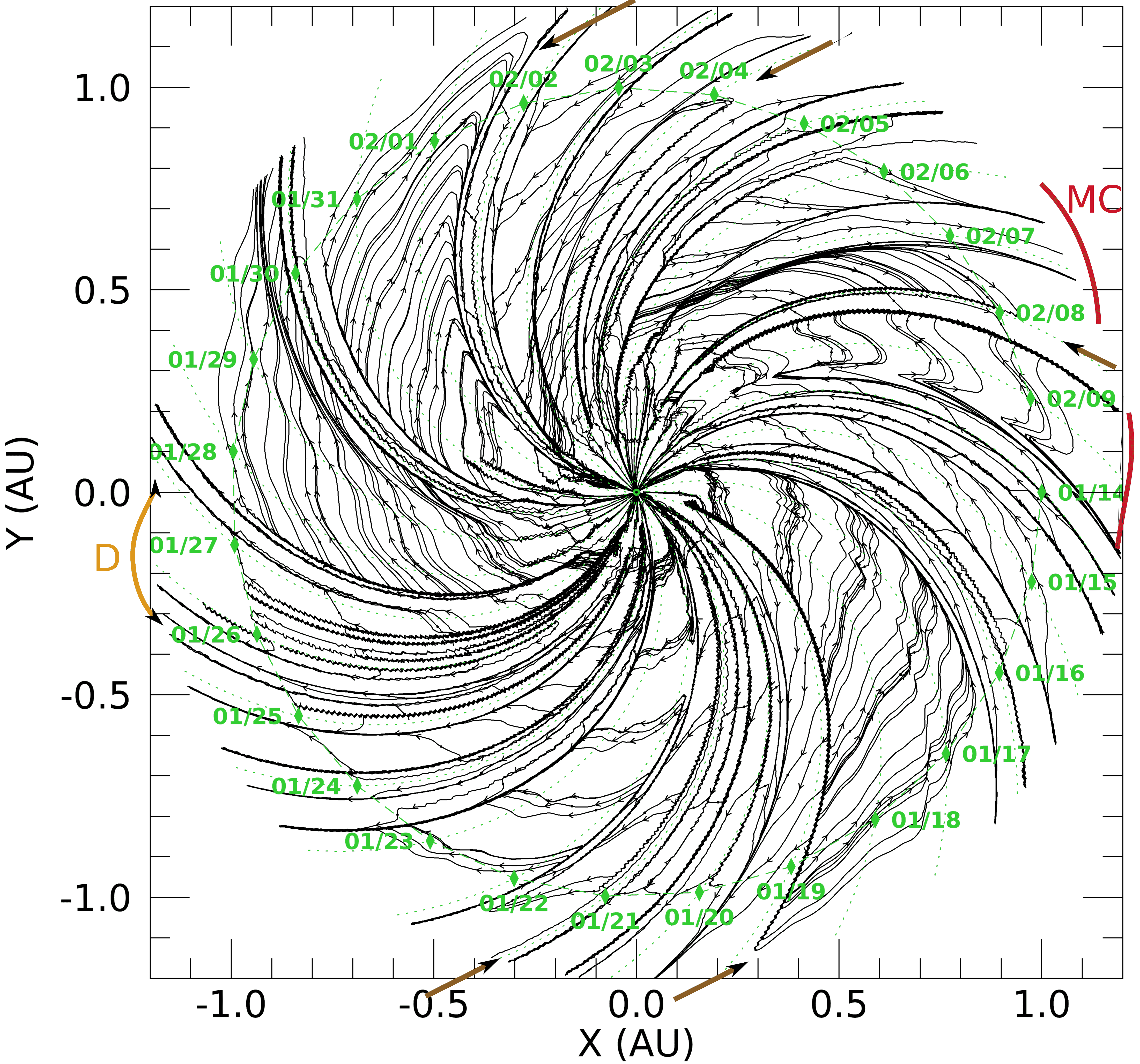}
  \end{minipage}
  \hfill
  \begin{minipage}[a]{0.7\textwidth}
    \includegraphics[width=\textwidth]{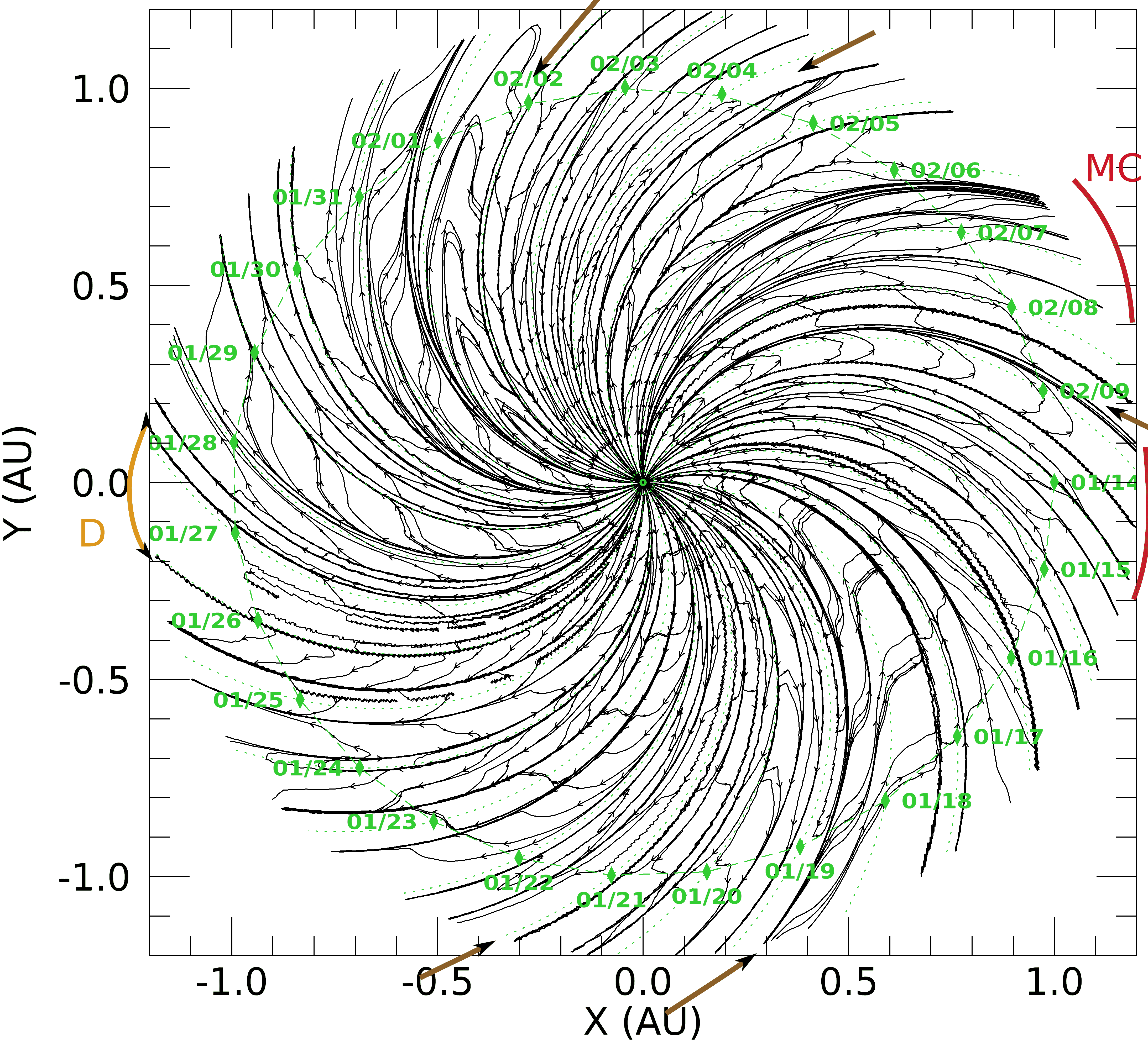}
    \caption{Predicted magnetic field line maps for the solar rotation period 14 January to 9 February 2014 (between CR 2145 and CR 2146) using the accelerating solar wind model of \citet{Tasnim2018} ({\it top}) and the constant solar wind model of \citet{Hagen2012} ({\it bottom}). Here D presents disconnected field lines and all the other symbols are as in Figures~\ref{mapping_1} and \ref{mapping_2}. 
}
\label{mapping_5}
\end{minipage}
\end{figure}

Figure~\ref{mapping_5} shows the maps for the accelerating and constant solar wind models (using {\it Wind} spacecraft data near $1$~AU data) 
between CR 2145 and CR 2146. The maps have strong similarities on a large scale, as found in Figures \ref{mapping_1} --\ref{mapping_3} for a period near solar minimum. 

Comparisons between the top and bottom panels of Figure~\ref{mapping_5} show that some differences exist that are very similar to those for Figures \ref{mapping_1} -- \ref{mapping_3}. For instance, field lines are more azimuthal for the accelerating solar wind model than the constant solar wind model. The same interpretation is adopted as for Figures \ref{mapping_1} -- \ref{mapping_3}.

Note that during this period two fast streams (marked as Fast stream 1 and Fast stream 2) briefly passed the {\it Wind} spacecraft, as did two ICMEs (the second ICME contains an MC region) recorded in the near-Earth CME list of Richardson and Cane [Revised June 08, 2018] \\ 
\url{http://www.srl.caltech.edu/ACE/ASC/DATA/level3/icmetable2.htm}, as noted by \citet{Li2016a}. The ICMEs occurred 8 - 9 February and the fast streams 14 - 16 and 22 - 24 January. The strong azimuthally oriented field lines are probably partly associated with the ICMEs and fast streams. 

\begin{figure}[!tbp]
  \centering
  \begin{minipage}[b]{0.8\textwidth}
    \includegraphics[width=\textwidth]{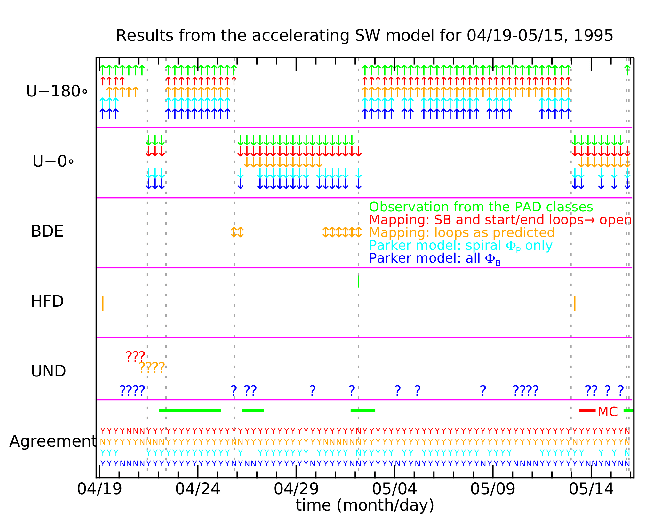}
  \end{minipage}
  \hfill
  \begin{minipage}[b]{0.8\textwidth}
    \includegraphics[width=\textwidth]{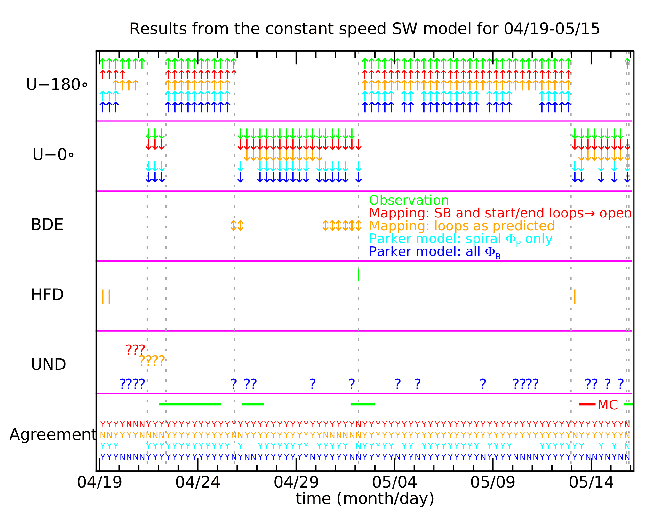}
    \caption{({\it Top}) Comparisons between the PAD classes observed ({\it green symbols}) and predicted using the magnetic field lines for the accelerating solar wind model ({\it red and orange symbols}) and Parker's spiral model ({\it blue symbols}) for the solar rotation 19 April to 15 May 1995 near solar minimum. {\it Arrows} and {\it question marks} correspond to definite versus undetermined PAD identifications. {\it Red (orange) symbols} correspond to artificial loops at sector boundaries and the start/ends of the period being interpreted as open (as predicted). Dark and light blue symbols correspond to values of $\phi_B$ within the Parker model's formal range and all four quadrants, respectively. {\it(Bottom)} Similar comparisons for the constant speed solar wind model \citep{Li2016a}. }
    \label{mapping_4}
  \end{minipage}
\end{figure}

\section{Comparisons between PAD Classes and the Predicted Magnetic Maps}
\label{PAD-classes}
Observations show that the pitch angle distributions (PADs) of suprathermal electrons are usually found in one of four distinct classes: (i) unidirectional strahls peaked at $0^\circ$ pitch \citep{Rosenbauer1977}, where the in-ecliptic angle $\phi_B=\tan^{-1}(B_\phi/B_r)$ the antisunward direction is defined by angles $90^\circ\leq\Phi_B\leq 270^\circ$ , (ii)  unidirectional strahls peaked at $180^\circ$ pitch \citep{Rosenbauer1977}, (iii) bi-directional electrons (BDE) with counterstreaming strahls, which result from being connected to the Sun along $\pm {\bf B}({\bf r})$ in a loop-like structure \citep{Gosling1987,Crooker2004a,Crooker2004b,Rouillard2010}, and (iv) heat-flux dropouts (HFD) with no observable strahls, which result from being disconnected from the Sun along $\pm {\bf B}({\bf r})$ \citep{McComas1989,Rouillard2010,Owens2013}. Knowing the magnetic connection to the Sun from the model magnetic maps,  we can predict the PAD classes for the suprathermal electrons as a function of time \citep{Li2016a}. This approach allows us to compare the predicted and observed PADs quantitatively. Apart from the $U-0^\circ$, $U-180^\circ$, BDE, and HFD classes defined above (in order), the acronym UND corresponds to situations when the PAD class is ``undetermined'', for  instance when the field line is incomplete or the observations are unclear. Here, $U$ defines the unidirectional strahls.


Consider first the period near solar minimum. Figure~\ref{mapping_4} compares the PAD classes observed by the {\it Wind} spacecraft (green arrows) with those predicted for the constant solar wind speed model, as reported by \citet{Li2016a}, and those predicted for the accelerating wind model. The PAD predictions with red and orange arrows correspond to the magnetic field maps with and without modifications (open field lines) at sector boundaries and at the start and end of the solar rotation period. Clearly the PAD classes predicted for the two solar wind models agree very well with each other and with the observations. 


\begin{center}
\begin{table}[!ht]
\small
\caption{ The PAD classes observed and predicted using the orientation and connectivity of magnetic fields lines for the accelerating wind model \citep{Tasnim2018} and the constant wind speed model \citep{Li2016a} for CR1895.}
\begin{tabular}{ |c|c|c|c|c|c|c| }
 \hline
  & U-$180^\circ$\ &U-$0^\circ$&BDE&HFD  &  UND & Agreement \\ [0.5ex] 
 \hline
\color{green} Observations   & \color{green} 51    &\color{green} 29&  \color{green}  0 &\color{green}1 &\color{green}0 & \color{green} - \\
 \hline
\hline
Accelerating wind & & & & & & \\
speed model & & & & & & \\
\hline
\color{red} Mapping: SBs and  &  \color{red} 48  & \color{red}31   &\color{red}0 &\color{red} 0& \color{red} 3& \color{red}94\%\\
\color{red} start/end loops$\rightarrow$open & & & & & & \\
 \hline
\color{orange} Mapping: loops as &\color{orange}46 & \color{orange}20&  \color{orange}8 & \color{orange}2&\color{orange}4 & \color{orange}80\%\\
\color{orange}predicted & & & & & & \\
 \hline
\color{cyan}Parker model:    &\color{cyan}38& \color{cyan}23&  \color{cyan}- &\color{cyan}- & \color{cyan}-&\color{cyan} 95\%\\
\color{cyan} spiral $\phi_{\rm p}$  only& & & & & & \\
 \hline
\color{blue}Parker model: all $\color{blue}\phi_{\rm p}$ &   \color{blue}38 & \color{blue}23&\color{blue}- &\color{blue}- & \color{blue} 20 & \color{blue}75\%\\
\hline
\hline
Constant wind  & & & & & & \\
speed model & & & & & & \\
 \hline
\color{red} Mapping: SBs and  &  \color{red} 47  & \color{red}31   &\color{red}0 &\color{red} 0& \color{red} 3& \color{red}95\%\\
\color{red} start/end loops$\rightarrow$open & & & & & & \\
 \hline
\color{orange} Mapping: loops as &\color{orange}45 & \color{orange}20&  \color{orange}8 & \color{orange}3&\color{orange}4 & \color{orange}79\%\\
\color{orange}predicted & & & & & & \\
 \hline
 \hline
\end{tabular}
\label{table1}
\end{table}
\end{center}

Table~\ref{table1} assesses statistically the results in Figure \ref{mapping_4} where we calculate the agreement by comparing the predicted PAD with the PAD observations. {\color{black} Descriptions of coloured arrows and symbols in Figure \ref{mapping_4} along with the associated agreement ratio in Table~\ref{table1} are as follows:
\begin{enumerate}
    \item Green arrows and symbols on Table~\ref{table1} show the observed PAD classes for CR 1895. The PAD classes in red are predicted by assuming the loops are open at  the start and end of the solar rotation period along with when they cross the SBs. Agreement is found in predicted $77$ (in red) PADs out of $81$ sampled PAD observations (in green) for the accelerating solar wind model, with a success rate of $95\%$. For the constant solar wind model $76$ of the $81$ samples agree ($94\%$).

\item Slightly worse agreement is found for both models using the unmodified field predictions (orange arrows). Agreement is found  in orange $65$ out of $81$ sampled PAD observations (in green) for the accelerating solar wind model, with a success rate of $80\%$ whereas success rate is $79\%$ for the constant solar wind model ($64$ (in orange) agree out of $81$ (in green) PAD observations).

\item \citet{Li2016a} also considered the agreement for magnetic field samples whose directions are within and outside the two quadrants allowed for the Parker solar wind model. The allowed quadrants correspond to azimuthal angles $\phi_{B} = {\rm \cos}^{-1} ({\bf B}\cdot{\bf r} / B r)$ in  the ranges $90 < \phi_{B} < 180^\circ$ and $270 < \phi_{B} < 360^\circ$. In these domains the accelerating wind model predicts unidirectional strahls for $95\%$ of the observed samples (cyan arrows in Figure \ref{mapping_4} where 58 cyan arrows agree with 61), very similar to the $94\%$ rate for the constant speed model \citep{Li2016a}.

\item However, the nominal Parker spiral model can not explain $20$ of the total $81$ samples since they have $\phi_{B}$  outside the two allowed quadrants. Therefore, if we consider all $81$ field samples, the Parker model only agrees with the observed PAD classes for $75\%$  (in blue) of the $81$ samples (in green), in contrast with the rates of $95\%$ and $94\%$ for the accelerating and constant speed models (Table~\ref{table1}), respectively.  
\end{enumerate} }
 

\begin{figure}[!ht]
  \centering
  \begin{minipage}[b]{1.0\textwidth}
    \includegraphics[width=0.8\textwidth]{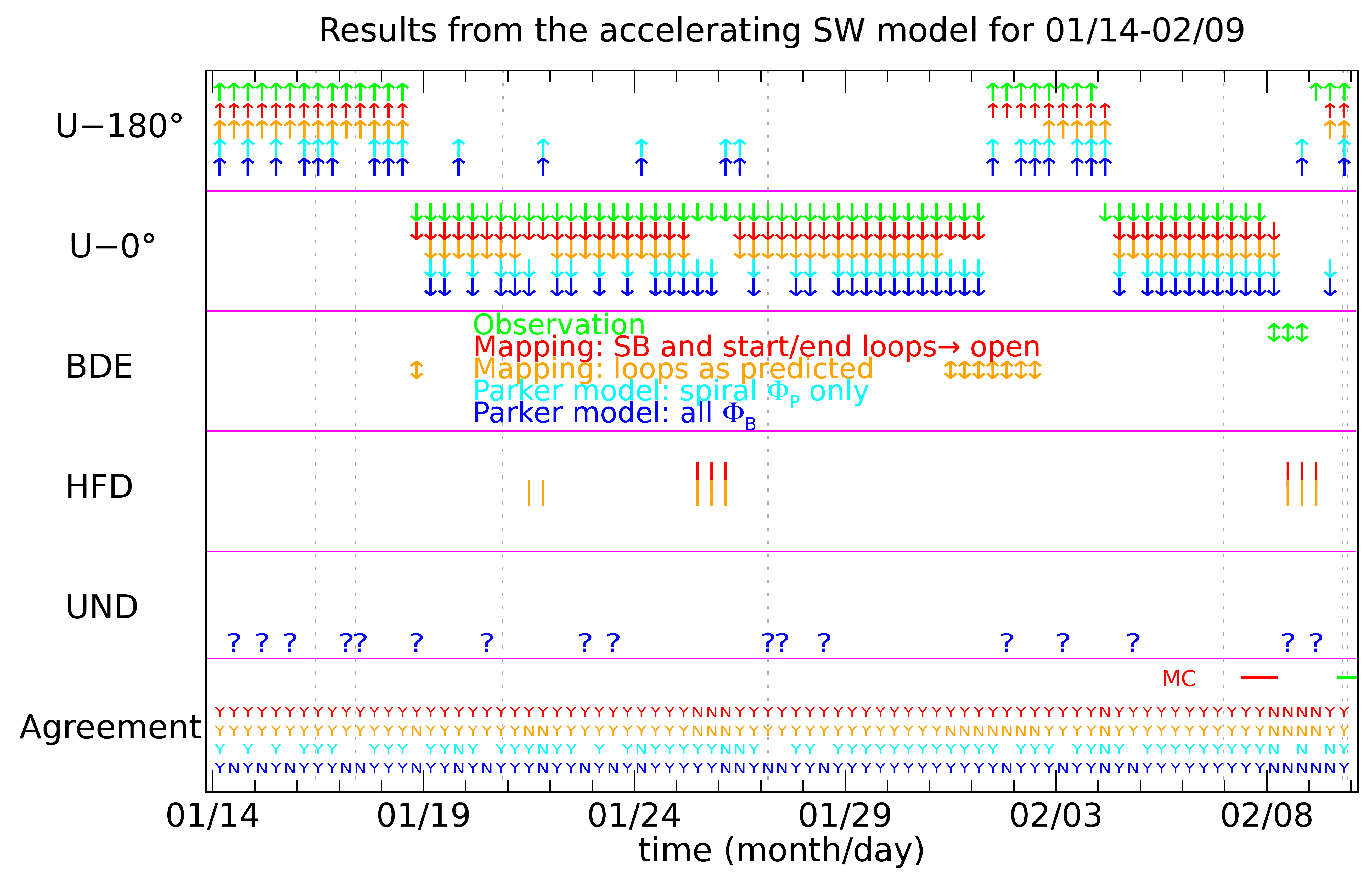}
  \end{minipage}
  \hfill
  \begin{minipage}[b]{1.0\textwidth}
    \includegraphics[width=0.8\textwidth]{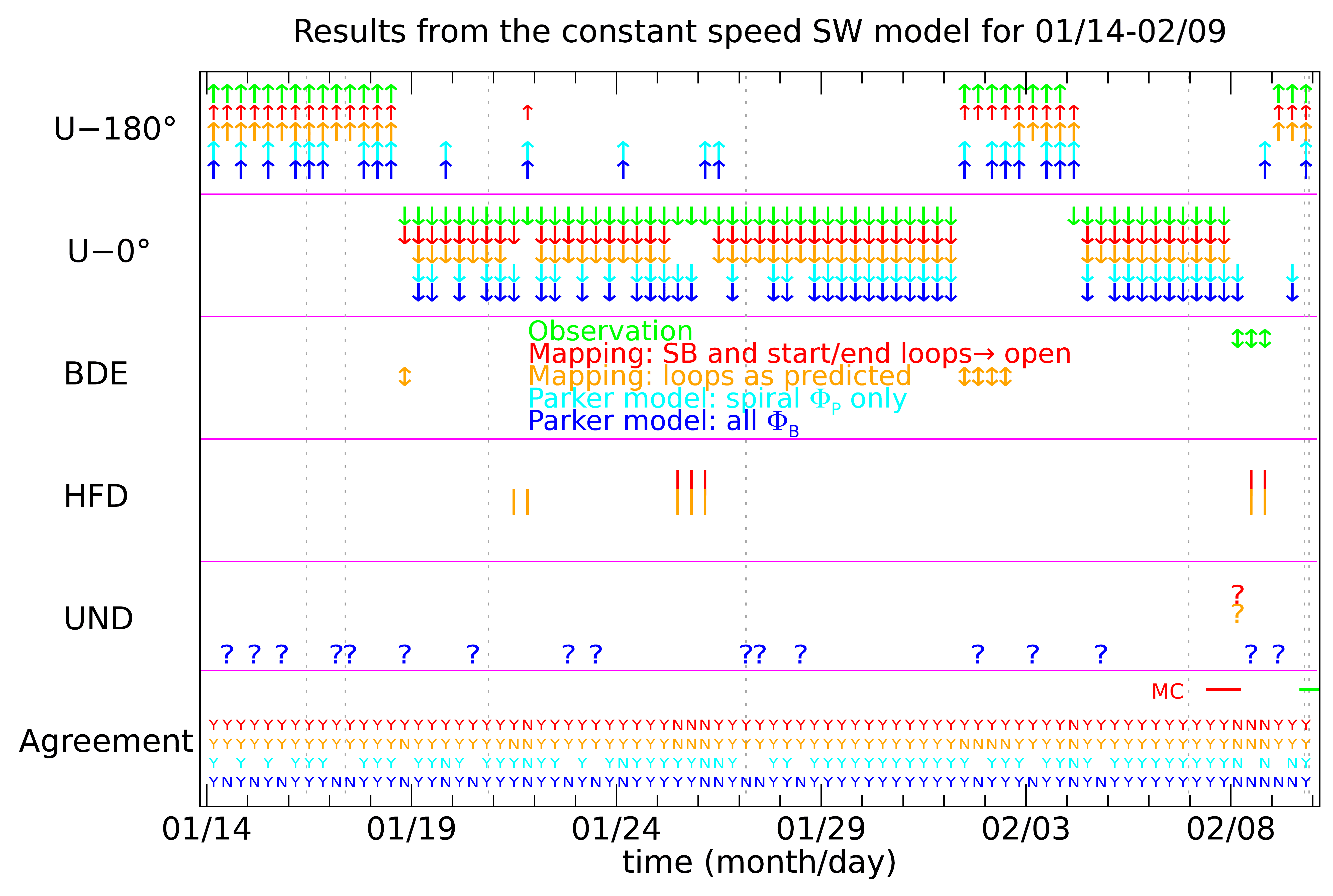}
    \caption{Comparisons between the observed and predicted for the ({\it top}) accelerating solar wind model and ({\it bottom}) constant wind speed model, as well as the Parker spiral model, for the period 14 January to 9 February 2014 near solar maximum. The format is as for Figure \ref{mapping_4}.}
    \label{mapping_6}
  \end{minipage}
\end{figure}

\begin{center}
\begin{table}[ht]
\small
\caption{Observed PAD classes and predicted using the accelerating wind model \citep{Tasnim2018} and the constant wind speed model \citep{Li2016a} for the period 14 January to 9 February 2014 near solar maximum.}
\begin{tabular}{ |c|c|c|c|c|c|c| }
 \hline
  & U-$180^\circ$\ &U-$0^\circ$&BDE&HFD  &  UND & Agreement \\
 \hline
\color{green} Observations   & \color{green} 25   &\color{green} 53&  \color{green}  3 &\color{green}0&\color{green}0 & \color{green}- \\
  \hline
\hline
Accelerating wind & & & & & & \\
speed model & & & & & & \\
\hline
\color{red} Mapping: SBs and  &  \color{red} 25  & \color{red}50   &\color{red}0 &\color{red} 6& \color{red} 0& \color{red}91\%\\
\color{red} start/end loops$\rightarrow$open & & & & & & \\
 \hline
\color{orange} Mapping: loops as &\color{orange}21 & \color{orange}44&  \color{orange}8 & \color{orange}8&\color{orange}0& \color{orange}81\%\\
\color{orange}predicted & & & & & & \\
 \hline
\color{cyan}Parker model:    &\color{cyan}23& \color{cyan}41&  \color{cyan}- &\color{cyan}- & \color{cyan}-&\color{cyan} 88\%\\
\color{cyan} spiral $\phi_{\rm p}$  only& & & & & & \\
 \hline
\color{blue}Parker model: all $\color{blue}\phi_{\rm p}$ &   \color{blue}23 & \color{blue}41&\color{blue}- &\color{blue}- & \color{blue} 17 & \color{blue}69\%\\
\hline
\hline
Constant wind  & & & & & & \\
speed model & & & & & & \\
\hline
\color{red} Mapping: SBs and  &  \color{red} 26  & \color{red}48   &\color{red}0 &\color{red} 6& \color{red} 1& \color{red}90\%\\
\color{red} start/end loops$\rightarrow$open & & & & & & \\
 \hline
\color{orange} Mapping: loops as &\color{orange}22 & \color{orange}46&  \color{orange}5 & \color{orange}7&\color{orange}1 & \color{orange}80\%\\
\color{orange}predicted & & & & & & \\
 \hline
 \hline
\end{tabular}
\label{table2}
\end{table}
\end{center}

Figure~\ref{mapping_6} presents observed and predicted PAD classes using {\it Wind} data, \citet{Tasnim2018}'s model, and \citet{Li2016a, Li2016b}'s model for the period 14 January to 9 February 2014 near solar maximum. Table~\ref{table2} assesses the  results statistically. The success rate for accelerating wind model during this solar rotation period is $91 \%$ if we assume the field lines are open at the sector boundaries and the start and end of the time interval (red symbols), while the corresponding predictions for the constant solar wind model has a success rate of $90 \%$. The other results of Table~\ref{table2} are very similar to those in Table~\ref{table1}. 

\section{Discussion and Conclusions}
\label{dis-con}
This paper develops a new algorithm to map magnetic field lines in the solar wind by solving Equation~\ref{runge-kutta} using the Runge-Kutta fourth-order method. This new algorithm allows us to assess the existing $\bm{B}$-step mapping algorithm developed by \citet{Li2016a, Li2016b}. The magnetic field line maps for these two algorithms are almost identical for both the accelerating and constant wind models for multiple time periods. This cross-validates the two mapping algorithms and allows us to confidently expect either algorithm to yield almost the same map for the same set of starting points. In addition, this paper improves the existing $\bm{B}$-step algorithm to provide unbiased and ``global" field lines, instead of maps that emphasize field lines near $1$~AU, by using a globally distributed set of starting points for the maps.

In this study, we have demonstrated, the generalization of the mapping approach of \citet{Li2016a, Li2016b} to use a more advanced solar wind model. This model includes acceleration of the solar wind, conservation of angular momentum, non-zero intrinsic azimuthal velocities at the inner boundary (nominally the photosphere) that allow a deviation from corotation, and non-zero azimuthal intrinsic magnetic fields at the inner boundary. 

We also shown that the mapped field lines (and so their directions and connectivities) are very similar on large scales (corresponding to several days near $1$~AU) for the two models, but with differences at smaller scales. Further the two models produce almost identical predictions for the PADs of superthermal electrons at $1$~AU, which agree with the observed PADS at the levels of $\approx 90 - 95\%$ for the two intervals considered, one near solar minimum and one near solar maximum. Moreover, the very similar results in Figures ~\ref{mapping_1} --  ~\ref{mapping_5} suggest that the comprehensive testing of \citet{Li2016a, Li2016b} against observational data for the constant wind speed model will apply with minimal changes to other predictions of the accelerating wind model. Thus, further work on mapping field lines and their inversions and connectivity from the corona to beyond $1$~AU can confidently use the new accelerating wind model \citep{Tasnim2018} instead of the simpler constant wind speed model \citep{Hagen2011, Hagen2012} used by  \citet{Li2016a, Li2016b}. It is worth emphasizing that significantly non-Parker spiral magnetic field lines are found for all the maps presented, for both wind models and for multiple solar rotations (both near solar maximum and solar minimum); this suggests that the assumption of Parker spiral field lines requires more care than often given and may often be inappropriate. 

While the two mapping approaches yield strongly similar maps on a broad scale for the two wind models, close inspection sometimes shows notable path and connectivity differences on small scales. See, for instance, the almost horizontal field lies near $1$~AU at the top of each map in Figures ~\ref{mapping_1} -- \ref{mapping_3}. These dissimilarities suggest that inclusion of acceleration of the radial wind speed, conservation of angular momentum,  and intrinsic non-radial, non-corotating velocities can have significant effects on magnetic field lines at small scales. This result agrees with the analytic expressions for the magnetic field components in the accelerating wind model, Equations (\ref{TCW-bphi}) - (\ref{Bphi-const}), due to the dependences of $B_{\phi}({\bf r}$) on the radial wind speed profile and the azimuthal velocity at the inner boundary.  

It thus appears that the new accelerating wind model can be confidently combined with the \citet{Li2016a,Li2016b,Li2016c} mapping approach, using either a {\textbf B} - step or Runge-Kutta algorithm as desired, to predict magnetic field maps and associated magnetic connectivities from the Sun to $1$~AU and beyond. These can be used to predict the PADs and time profiles of superthermal electrons and SEPs, whether from slowly-evolving solar structures or from flares and moving shocks. Comparisons with observations near $1$~AU or  from the {\it Parker Solar Probe}, {\it Messenger}, {\it Beppi-Colombo}, or future {\it Solar Orbiter} closer to the Sun will test the magnetic field maps, solar wind models, and models for particle propagation, acceleration, and scattering. They may provide evidence for magnetic field evolution and solar wind physics not included in the models. { \color{black} In addition, magnetic maps can be initialised using results from \citet{Veselovsky2006}, and then the predicted field lines can be compared with the predictions from this paper.} An interesting point is that the ${\bf v}({\bf r},t)$ and ${\bf B}({\bf r},t)$ data required to initialise the solar wind models and provide global maps need not be at $1$~AU. Comparisons between the maps predicted with data from {\it Parker Solar Probe}, {\it Messenger}, {\it Beppi-Colombo}, and {\it Solar Orbiter} close to or inside $0.3$3~AU with those for spacecraft data near $1$~AU may well be particularly useful in assessing the solar wind models and evidence for evolution of the plasma, magnetic field, and SEPs between the Sun and $1$~AU. 

\section{Acknowledgements}

\small{\rmfamily{We would like to thank CDWeb of NASA for the Wind data. I acknowledge the financial support of Prof. Gary P. Zank from The Center for Space Plasma and Aeronomic Research (CSPAR), the University of Alabama in Huntsville (UAH), and continuous guidance of Prof. Iver H. Cairns from the school of physics, the University of Sydney throughout this research.}} \\

\section*{Disclosure of Potential Conflict of Interest}  
The authors declare that they have no conflicts of interest.




\bibliographystyle{spr-mp-sola}

\tracingmacros=2
\bibliography{SWbib}  

\IfFileExists{\jobname.bbl}{} {\typeout{}
\typeout{****************************************************}
\typeout{****************************************************}
\typeout{** Please run "bibtex \jobname" to obtain} \typeout{**
the bibliography and then re-run LaTeX} \typeout{** twice to fix
the references !}
\typeout{****************************************************}
\typeout{****************************************************}
\typeout{}}

\end{article} 

\end{document}